\documentclass[twocolumn,showpacs,pra,floatfix]{revtex4}
\usepackage{amsmath}
\usepackage{amssymb}
\usepackage{amsfonts}
\usepackage{delarray}
\usepackage{graphicx}
\usepackage{float}
\usepackage{subfigure}
\usepackage{longtable}
\newcommand{\bra}[1]{\langle #1|}
\newcommand{\ket}[1]{|#1\rangle}
\newcommand{\braket}[2]{\langle #1|#2\rangle}
\newcommand{\Tr}{\mathrm{Tr}}

\begin{document}

\title{Quantum decoherence without reduced dynamics}
\author{P. W.~Bryant}
\affiliation{Center for Complex Quantum Systems,
             Department of Physics,
             University of Texas at Austin, Austin, Texas 78712}

\date{\today}

\begin{abstract}
With a choice of boundary conditions for solutions of the
Schr\"odinger equation, state vectors and density operators even for
closed systems evolve asymmetrically in time.
For open systems, standard quantum mechanics consequently
predicts irreversibility and signatures of the extrinsic arrow of time.
The result is a new framework for the treatment of decoherence,
not based on a reduced dynamics or a master equation.
As an application, using a general model we quantitatively match
previously puzzling experimental results and can conclude that they
are the measurable consequence of the indistinguishability of 
separate, uncontrolled interactions between systems and their
environment.
\end{abstract}

\maketitle

\section{Introduction}
One cannot overstate
the practical importance of understanding quantum decoherence.
The remarkable emergence of quantum engineering and the pursuit of quantum
computation have resulted in the creation of entire
industries relying upon the minimization of the effects of decoherence.

Less practical but perhaps more significant is the insight to be gained into
the theory of fundamental processes.
Many have sought a natural explanation for this experimental signature of
an arrow of time for quantum 
systems, which have states represented theoretically by 
density operators with intrinsically reversible evolution in
time~\cite{wheeler_zurek_quantum_1983,schulman_times_1997,zeh_physical_2007}.

Phenomenologically, one imagines irreversible behavior to result from 
uncontrolled interactions between an experimental system
and its environment.
An arrow of time is thus usually accommodated in quantum theory
by modeling the system of interest as a subsystem embedded
within a quantum mechanical 
environment~\cite{von_neumann_meas_irrevers,petruccione_open_quantum_systems}.
While the time evolution of the environment plus subsystem is unitary,
the time evolution of the subsystem alone need not be.
In practice, one typically neglects memory effects of the
environment~\cite{kossakowski_quantum_1972,lindblad_generators_1976}
and models this reduced dynamics of the subsystem using a
Lindblad-Kossakowski master equation.

Nature has provided us also with decaying states and scattering
resonances, which are thought to be signatures of an asymmetry in 
time~\cite{fortschritte_der_physik_1} even for closed systems.
This time asymmetry is called \emph{intrinsic} 
to distinguish it from irreversibility and
the extrinsic arrow of time, which is consequent to environmental interactions.
It is endowed at the microphysical level and is thus 
understood to have a different physical origin~\cite{bohm_harshman_book_1998}.
If one demands a rigorous, theoretical unification of
scattering resonances with decaying states, then one requires time evolution 
generated by a semigroup~\cite{physica_a_1997}.
Semigroup time evolution is a theoretical expression
of microphysical time asymmetry.

Kossakowski and Rebolledo have recently begun to apply
concepts from the intrinsically time asymmetric theory to the formulation of
a new master equation~\cite{kossakowski_nonmarkovian_2007}.
Their approach is based on the similarities between the energy spectra of subsystems
embedded in an environment and of scattering resonances embedded in a continuum.

Here we take a different approach.
There is a direct
relationship between the intrinsic time asymmetry and the extrinsic arrow of time.
If one chooses microphysically asymmetric time evolution, one must very carefully
construct the theoretical images of quantum systems.
One also finds a new correspondence between time evolution parameters and the
passage of time in the physical universe.
It is just this new correspondence, and not some mathematical effect,
that leads automatically to predictions of decoherence for open systems.

After explaining briefly in Section \ref{sec:boundary_conditions}
what we mean by intrinsically asymmetric time evolution,
we will demonstrate in Section \ref{sec:consequences}
how one must represent experimental systems.

In Section \ref{sec:ap} is an application.
We derive a predictive probability
for the measured decoherence of quantum systems undergoing Rabi oscillations.
Furthermore, we match a general experimental result that has been particularly puzzling.
Using our new approach, we conclude that what has been called
Excitation Induced Dephasing~\cite{ramsay_prl_2010}
is a consequence of
the indistinguishability of separate, uncontrolled interactions between quantum
systems and their environment.

\section{Groups and Semigroups}
\label{sec:boundary_conditions}
The theoretical image of a quantum mechanical system is an 
operator algebra defined in a linear scalar-product space,
$\Phi$~\cite{bohm_textbook}.
The vectors $\phi_n$ span the space $\Phi$, and every linear combination
of the $\phi_n$ can represent the state of the physical system.
For simplicity, in this section let us 
consider a single vector, $\phi$, that satisfies the Schr\"odinger equation.
The vector $\phi$ spans a one-dimensional subspace of $\Phi$, and this
ray represents the quantum system in a pure state.

For the scalar-product space, $\Phi$, one has historically chosen
the Hilbert space: $\Phi=\mathcal{H}$.
With this choice, 
Stone and von Neumann showed~\cite{stone_1932,von_neumann_1932} that
the time evolution of solutions is given generally by
\begin{equation}
\label{state_evolution}
\phi(t)=e^{-\frac{i H t}{\hbar}}\phi\,,\qquad-\infty<t<\infty.
\end{equation}
To require $\phi\in\mathcal{H}$ is to constrain $\phi$, just as one
constrains the general solutions of any differential equation by enforcing
proper behavior at the boundaries.
For this reason, one calls $\phi\in\mathcal{H}$ a choice of boundary conditions.

The time evolution in \eqref{state_evolution} can
be described by the one-parameter group of unitary operators
\begin{equation}
\label{state_group}
U(t)=e^{-\frac{i H t}{\hbar}}\,,\qquad -\infty<t<\infty.
\end{equation}
For every evolution, $U(t)$, there exists the inverse, $U(t)^{-1}$,
given by $U(t)^{-1}=U(-t)$.
This means that
$U(-t)\phi$ is the time-reversed version of $U(t)\phi$, and that
\emph{both are solutions of the Schr\"odinger equation}.
The resulting dynamics is intrinsically symmetric in time.

One often chooses state vectors from a different scalar-product
space~\cite{dirac_qm_book},
though usually without realizing it.
When using Dirac kets ($\ket{E}$, $\ket{\vec{x}}$, etc.) with continuous spectra 
of eigenvalues,
one has chosen $\phi\in\mathcal{S}\subset \mathcal{H}$,
where the Schwartz space, $\mathcal{S}$, is a
subset of $\mathcal{H}$~\cite{bohm_textbook}.

Though they are measured differently, one typically assumes that scattering
resonances are physically equivalent to decaying states.
To find a rigorous theory in which the two can be considered also
\textit{mathematically} equivalent,
one must constrain the state vectors even further~\cite{bohm_bryant_quantal_2008}:
\begin{equation}
\label{eq:hardy_axiom_1}
\textrm{set of possible states }\{\phi\} = \Phi_-\subset\mathcal{H}\subset \Phi^\times_-,
\end{equation}
where $\Phi_-$ denotes the Hardy space of the lower complex
semiplane, and $\Phi^\times_-$ is its dual.
Note that, in practice, \eqref{eq:hardy_axiom_1} is not a limiting
restriction.
Any vector in the Hilbert space can be approximated 
with arbitrary precision by vectors in the Hardy
space~\cite{bohm_monograph_1989}.
This section contains only a sketch, and for details the interested reader is 
referred to~\cite{bohm_annals_of_physics_small_preprint,
gadella_hardy_rhs,fortschritte_der_physik_1} and 
and the numerous references therein.

The only mathematical consequence of \eqref{eq:hardy_axiom_1} we shall
consider here is that solutions of the Schr\"odinger equation
become~\cite{bohm_annals_of_physics_small_preprint}
\begin{equation}
\label{asym_states}
\phi(t)=e^{-\frac{i H t}{\hbar}}\phi\,,\qquad 0\leq t<\infty.
\end{equation}
Note the lower bound on the time parameter.
This is no longer time evolution given by the unitary group of \eqref{state_group}.
The time evolution is instead given by the 
one-parameter semigroup of operators,
\begin{equation}
\label{semigroup_states}
U(t)=e^{-\frac{i H t}{\hbar}}\,,\qquad 0\leq t<\infty.
\end{equation}
Being a semigroup means that the inverse, $U(t)^{-1}$, of any element,
$U(t)$ with $t>0$, does not exist.
In other words, \emph{the time-reversed version of $U(t)\phi$ is
no longer an available solution of the Schr\"odinger equation}.
This dynamics is intrinsically asymmetric in time~\cite{bohm_harshman_book_1998}.
From a choice of boundary conditions,
asymmetric time evolution is endowed at the microphysical level 
and is independent of any interaction of a system with its environment.

The use of semigroups is not new to physics.
First, expert readers will notice that \eqref{semigroup_states} is \textbf{not}
the semigroup often assumed for the family of dynamical maps for a
reduced density matrix~\cite{petruccione_open_quantum_systems}.
It has also been suggested in the past
that one might describe relativistic, unstable particles using
representations of the Poincar\'e 
semigroup~\cite{schulman_unstable_poincare_1970,comi_poincare_1975,
alicki_poincare_1986,exner_poincare_1983}.

\section{Appearance of Decoherence}
\label{sec:consequences}
With our choice of boundary conditions \eqref{eq:hardy_axiom_1},
and thus the microphysical time asymmetry, comes a handful
of consequences affecting 
the representation of experimental systems by their theoretical images.
For open systems,
predictions of decoherence follow automatically from
standard quantum mechanics.

\subsection{Coordinates and Parameters}
In a recent review of time in quantum mechanics~\cite{zeh_compendium_2009},
Zeh explains that, for the non-relativistic theory,
one identifies the parameter $t$ in the Schr\"odinger equation
with ``Newton's absolute,'' or coordinate, time.
Because in the master equation formalism, one models experimental systems
as subsystems of arbitrarily large environmental systems also evolving in time,
consistency requires one to identify $t$ as such.

In some cases, however, it is known that such an identification
can be problematic.
For instance, to formulate a sensible version of the time-energy uncertainty
relation, one must distinguish between ``external time,'' which is
measured by laboratory clocks, and ``intrinsic time,'' which parametrizes
the dynamical evolution of experimental systems~\cite{busch_timeenergy_2008}.

We will demonstrate that the correct
application of the intrinsically time asymmetric theory \emph{requires}
one to distinguish between external and intrinsic time.
What is called external time in~\cite{busch_timeenergy_2008} we will
represent with \textbf{time coordinates}.
Time coordinates are the time part of the space-time coordinates.
They are physically insignificant time labels for events, and
their values are given by the laboratory clocks, which are not
dynamically connected to experimental systems.
We will label time coordinates with a tilde: $\tilde{t}$.

What is called intrinsic time in~\cite{busch_timeenergy_2008} we must
identify with the $t$ in the Schr\"odinger equation and its
solutions~\eqref{asym_states}.
We will represent this intrinsic time with \textbf{time parameters},
which parametrize the time evolution of state vectors.
Time parameters always correspond to durations, and we will continue
to label them with the letter $t$.
In Table~\ref{tab:copar} is a reminder of the distinction we will have to make.
\begin{table}[t]
\begin{center}
\begin{tabular}{|l|c|l|}
\hline
\textbf{Name} & \textbf{Symbol} & \textbf{Description} \\
\hline
coordinate & $\tilde{t}$ & 
$\begin{array}{l}
\textrm{time labels for events;} \\
\textrm{marked by clocks}
\end{array}$ \\
\hline
$\begin{array}{l}
\textrm{semigroup} \\
\textrm{time parameter}
\end{array}$ & $t$ & 
$\begin{array}{l}
\textrm{parameter for Schr\"odinger} \\
\textrm{equation and solutions } \eqref{asym_states} 
\end{array}$\\
\hline
\end{tabular}
\caption{\label{tab:copar}Time coordinates and parameters.}
\end{center}
\vspace{-0.6cm}
\end{table}

Considering the time asymmetric boundary conditions,
that one must distinguish carefully between coordinates and parameters is
obvious because possible values of $t$ and $\tilde{t}$ are no longer chosen from the
same interval:
\begin{equation}
\label{eq:domains}
t\in [0,\infty)\qquad\textrm{and}\qquad\tilde{t}\in (-\infty,\infty).
\end{equation}

In what follows, we will assume that one can infer from measurement or 
from a preparation
procedure the nature of a density operator meant to represent the state of
a physical system.
Measurements and preparations
are performed in the coordinate time, $\tilde{t}$, of the laboratory,
so this inferred density operator is also a function of $\tilde{t}$.
Because only time durations are significant, we will call it
$\rho(\tilde{t}-\tilde{t}_{prep})$, where $\tilde{t}_{prep}$ is the time coordinate
value when the state of the system is prepared (see below.)

Any density operator written as $\rho(t)$
is understood to be a function of the time parameter, $t$, of theory.
It is calculated using standard quantum mechanics.
And as explained below, for a given system one cannot always equate
$\rho(\tilde{t}-\tilde{t}_{prep})$ with $\rho(t)$.
To summarize,

\begin{center}
\begin{tabular}{ccc}
\textbf{Calculated} & $\qquad$ & \textbf{Inferred from measurement} \\
$\rho(t)$ &  & $\rho(\tilde{t}-\tilde{t}_{prep})$ \\
\end{tabular}
\end{center}

\subsection{Preparation Time and $t=0$}
In the theory, there is now a distinguished value, $t=0$, of the
parameter used in \eqref{asym_states} to 
parametrize the time evolution of state vectors.
This distinguished value of time is
a mathematical feature that is phenomenologically significant and
identifiable in comparisons with experimental data.

One defines a physical observable by a prescription for how
it is to be measured.
In the theory, one represents observables with linear operators in
a space $\Phi$~\cite{bohm_textbook}.
In the Schr\"odinger picture,
the Born probability to find the observable $\Lambda=\ket{\psi}\bra{\psi}$ in 
the state $\rho(t)=\ket{\phi(t)}\bra{\phi(t)}$ is
\begin{equation}
\label{asymmetric_born}
\mathcal{P}_{\Lambda}\big(\rho(t)\big)=\Tr\big(\Lambda\,\rho(t)\big)
=|\braket{\psi}{\phi(t)}|^2,
\qquad 0\leq t<\infty.
\end{equation}
Here the time evolution comes from \eqref{asym_states},
and the calculated probability compared with experimental measurements
exists only for $0\leq t<\infty$.

For any experimental system present in a laboratory, there is also a special time:
the coordinate value of the preparation time, $\tilde{t}_{prep}$.
It is the time on the clock at which the state of that system
has been prepared such that it is representable by, say, $\rho(t)$.
It is also the time after which a detector can possibly register an 
observable~\cite{bohm_time_asym_pra_1999}:
\begin{eqnarray}
\label{eq:obs_prep_meas}
&\textrm{the observable represented by }\Lambda \nonumber \\
&\textrm{is registered
at }\tilde{t}\geq \tilde{t}_{prep}.
\end{eqnarray}

Comparing \eqref{eq:obs_prep_meas} with \eqref{asymmetric_born},
one identifies
the semigroup time parameter value $t=0$ of the theory with the time coordinate
corresponding to the preparation of the state of an experimental system,
$\tilde{t}_{prep}$:
\begin{center}
\begin{tabular}{ccc}
\textbf{Theory} & & \textbf{Physical Identification} \\
$t=0$ & $\Leftrightarrow$ & $\begin{array}{c}\textrm{preparation time of the state}\\
                                   \textrm{of the experimental system}\end{array}$ \\
\end{tabular}
\end{center}
For non-relativistic applications, then, the relation between coordinates
and parameters is trivial.
For a measurement performed at the duration $\tilde{t}-\tilde{t}_{prep}$ after
preparation, one has
\begin{equation}
t \Leftrightarrow \tilde{t}-\tilde{t}_{prep}.
\end{equation}
By causality, $\tilde{t}\geq\tilde{t}_{prep}$, and one cannot
contradict what is in \eqref{eq:domains}.

\subsection{Experimental Ensembles}
\label{sec:ensembles}
The results of quantum mechanical calculations are probabilities meant
to be compared with averages over many identical measurements performed
on identically prepared, experimental systems~\cite{dirac_qm_book}.
For actual experiments, of course, one replaces the notion of ``identical''
with something like ``as similar as possible.''
By \textbf{ensemble}, then,
we refer to the collection
of such identically prepared, experimental systems meant to undergo measurement.
This is the usual understanding of ensembles, but we emphasize
that we have no comment on the 
\textit{Ensemble Interpretation}~\cite{ballentine_statistical_1970} of
quantum mechanics.
Nor will we reference another level of abstraction in which
one imagines an ensemble of state vectors~\cite{pearle_ensemble_vectors_1989}
or of wave functions~\cite{zeh_physical_2007}.

While a density operator represents in principle the state of a single member 
as well as the state of 
the entire ensemble, a measurement on just one member alone is useless.
An ensemble always exists for any quantitatively useful experiment.

Because of our identification of the parameter value $t=0$ with
the time coordinate corresponding to the preparation of the state of
an experimental ensemble,
if $\rho(t)$ is to represent the state of all members of an
experimental ensemble, then we require the following rule:
\begin{description}
\item[Rule for the application of theory]$\,$\\
Every member of the ensemble,
at the moment it is prepared such that it is representable
by the density operator $\rho(t)$, is represented by that operator at the 
parametric time zero, $\rho(t=0)$,
\textit{regardless of the coordinate time in the lab at which the preparation occurs}.
\end{description}

This rule is already followed in the reduction of experimental data.
This is especially clear in dynamical experiments
performed repeatedly on single members of an
ensemble~\cite{nagourney_dehmelt_shelved_1986,
bergquist_qjumps_1986,sauter_toschek_qjumps_1986,peik_qjumps_1994,
meekhof_rabi_1996,brune_rabi_1996,petta_coherent_2005}.
Because the rule is followed when reducing data, its appearance in the theory
is quite natural.

\subsection{Theoretical Image of an Open System}
\begin{figure}[ht]
\includegraphics[trim = 4cm 19.2cm 6cm 5.35cm clip, width=.5\textwidth]{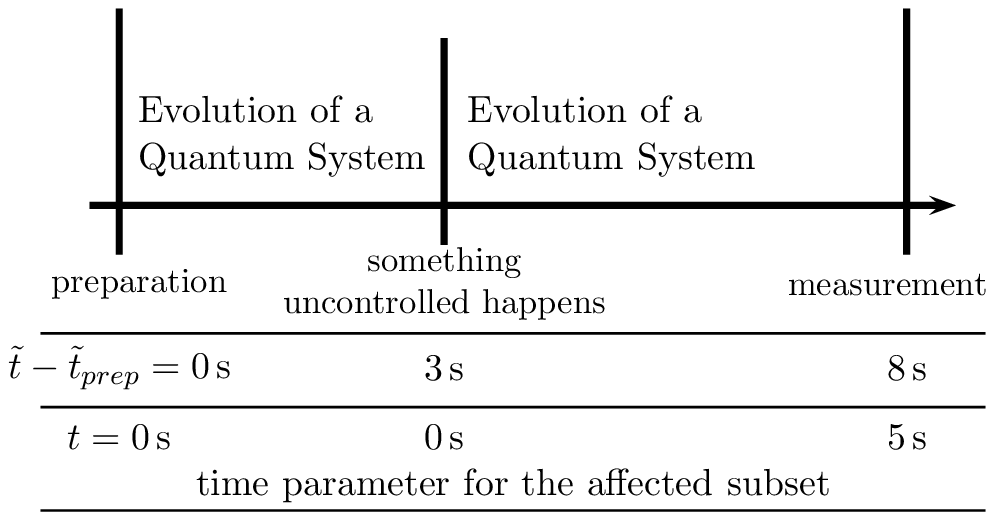}
\caption{Schematic timeline for an open quantum system.}
\label{fig:timeline}
\end{figure}

To understand why predictions of decoherence follow for open systems,
consider a hypothetical sequence of three events pictured in the
timeline in Figure~\ref{fig:timeline}:
\begin{enumerate}
\item A physicist prepares the state of $N$ members of an experimental
ensemble at,
according to the laboratory's clock, $\tilde{t}_{prep}$.
Depending on the experiment, of course, individual members of the ensemble may have
different values of $\tilde{t}_{prep}$.
In some experiments, these different values are recorded as ``time stamps.''
\item After a duration of $\tilde{t}-\tilde{t}_{prep}=3\,\textrm{s}$, something
uncontrolled (and not directly known) perturbs some members of the ensemble.
\item After the uncontrolled occurrence, the system evolves undisturbed until
finally, at $\tilde{t}-\tilde{t}_{prep}=8\,\textrm{s}$, the physicist
performs an active measurement.
\end{enumerate}
Because of the uncontrolled occurrence in Figure~\ref{fig:timeline},
the system is called an open system.
One can always generalize it, but for our purposes, this very simple example suffices.

\begin{figure}[hb]
\includegraphics[trim = 4.5cm 17.75cm 6cm 8.8cm, clip, width=.5\textwidth]{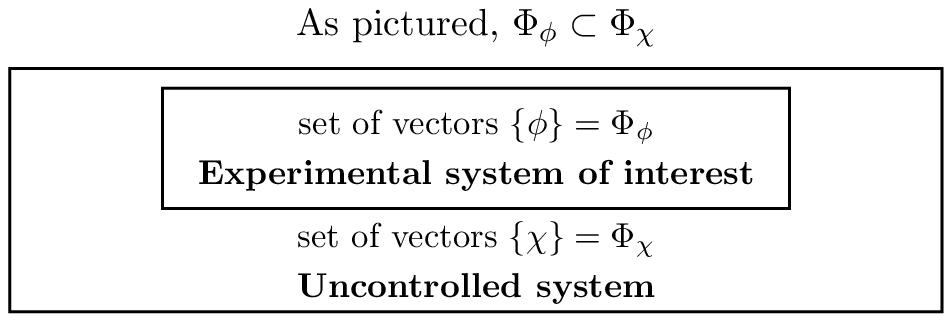}
\caption{The extent of two closed systems.}
\label{fig:systems}
\end{figure}

To construct the theoretical image of this sequence, standard quantum mechanics requires
the two closed systems pictured in Figure~\ref{fig:systems}.
The vectors $\phi_n$ span the space $\Phi_\phi$, and they can describe
the state of the experimental system of interest.
The vectors $\chi_n$ span $\Phi_\chi$, and they can describe the
state of the uncontrolled system.
By definition, any density operator defined in $\Phi_\phi$ can
describe the state of the system of interest and of \emph{nothing external
to it}.

Assume that the physicist prepares the ensemble of size $N$
to be in a pure state.
In the theory,
one then represents the state of the ensemble with an operator projecting
into a one-dimensional subspace of $\Phi_\phi$ spanned by, say $\phi_1$.
Before any uncontrolled experimental occurrence, i.e., for
$\tilde{t}-\tilde{t}_{prep} < 3\,\textrm{s}$ in Figure~\ref{fig:timeline},
the appropriate density operator at $\tilde{t}$ is
\begin{equation}
\label{eq:pure_op}
\rho(\tilde{t}-\tilde{t}_{prep})\Leftrightarrow\rho_\phi(t)
=U(t)\ket{\phi_1}\bra{\phi_1}U^\dagger(t).
\end{equation}
In \eqref{eq:pure_op}, the identification $t\Leftrightarrow\tilde{t}-\tilde{t}_{prep}$
is implicit.

If the uncontrolled event in Figure~\ref{fig:timeline} does not coherently
affect the entire ensemble, it affects only a subset of size $m$.
And by definition, the uncontrolled occurrence cannot be described
by the (closed) dynamics of the system of interest.
The state of the $m$ affected members of the ensemble
simply ceases to be representable by $\rho_\phi(t)$.
Instead, one must represent their state with 
a new operator, $\rho_\chi(t)$, which is in general defined in 
a different space, $\Phi_\chi$.
And according to the rule above, one must initially represent
the state of the affected subset with $\rho_\chi(t=0)$.
By preparing (passively) the state anew,
the uncontrolled event resets the time parameter, $t$.
This is noted in the bottom of Figure~\ref{fig:timeline}.

The state of the unaffected $N-m$ members remains representable by $\rho_\phi(t)$.
For $\tilde{t}-\tilde{t}_{prep} \geq 3\,\textrm{s}$,
\begin{equation}
\label{eq:mixed_op}
\rho(\tilde{t}-\tilde{t}_{prep})\Leftrightarrow a_1\,\,\rho_\phi(t_1) + a_2\,\,\rho_\chi(t_2).
\end{equation}
In \eqref{eq:mixed_op}, 
$a_1$ and $a_2$ are weights that will depend on $N$ and $m$,
as well as on the nature of the physical systems.

Note that $\rho_\phi(t)=\rho_\chi(t)$ is possible but certainly not required.
And even if they are equal, $\rho(\tilde{t}-\tilde{t}_{prep})$ in \eqref{eq:mixed_op}
will represent a mixed state because $t_1\neq t_2$ for any given value of the 
coordinate $\tilde{t}$ (see Figure~\ref{fig:timeline}.)

If $\Phi_\phi\neq\Phi_\chi$, then \eqref{eq:mixed_op} may not be well defined.
To treat it correctly one must eventually define how an operator
defined on $\Phi_\phi$ might be extended or limited to the space $\Phi_\chi$.
When $\Phi_\chi$ represents an uncontrolled environment, having
$\Phi_\phi\neq\Phi_\chi$ will generally lead to predictions of dissipation.
Further discussion can be left for another paper because, for the 
experiments we analyze in Section~\ref{sec:ap}, from the data
we will deduce that, immediately after the uncontrolled occurrence, $\Phi_\phi=\Phi_\chi$.

Now consider the time parameter for the affected subset of the ensemble.
At measurement
\begin{equation}
\label{eq:time_values}
8\,\textrm{s}=\tilde{t}-\tilde{t}_{prep}\neq t_2=5\,\textrm{s}.
\end{equation}
To conclude \eqref{eq:time_values},
we have made use only of the definition of open quantum systems
and of the intrinsic time asymmetry;
our treatment of
this hypothetical sequence is clearly independent of any model.
\emph{The time parameter values, $t$, do not
correspond simply to the passage of absolute time
in the physical universe.}

If the process of moving from \eqref{eq:pure_op} to \eqref{eq:mixed_op} can be
called ``branching'' of the density operator, then after multiple instances of 
environmental interference one has a density operator with many branches:
\begin{equation}
\label{eq:branched}
\rho(\tilde{t}-\tilde{t}_{prep})\Leftrightarrow a_1\,\,\rho_\phi(t_1) + a_2\,\,\rho_\chi(t_2)
+ \ldots \equiv \rho(\{t\}),
\end{equation}
where $\{t\}$ is a \emph{set} of time parameters.

A density operator for a pure state always 
branches into a density operator for a mixed state.
This is a prediction of decoherence, which is an experimental signature
of the quantum mechanical arrow of time.
And though this branching is for the theory very simple,
given our choice of boundary conditions it follows necessarily
from the correct application of standard quantum mechanics.

One could conceivably develop this approach to decoherence
without using the semigroup at all.
Based upon the remarkable experiments in which single
members of an experimental ensemble are repeatedly prepared and
measured~\cite{nagourney_dehmelt_shelved_1986,
bergquist_qjumps_1986,sauter_toschek_qjumps_1986,peik_qjumps_1994,
meekhof_rabi_1996,brune_rabi_1996,petta_coherent_2005},
from physical considerations one must already 
reset the time parameter to $t=0$ whenever ensemble members
are actively prepared anew.
But we have included the full development here because
we enjoy how time asymmetric boundary conditions~\eqref{eq:hardy_axiom_1}
lead automatically to the experimental signatures of irreversibility.

\section{Application:  Rabi Oscillations Experiments}
\label{sec:ap}
As a test of our approach to open quantum systems, let us now analyze
actual experimental systems within the framework of microphysical
time asymmetry and the branching density operators.
Our approach will allow us to address the indistinguishability of
experimental ensemble members, and as a consequence
we will find that our model compares very well with measurements.

\subsection{Rabi Oscillations}
The theory of Rabi oscillations is well established~\cite{dodd_atoms_1991}.
Consider a two level system, with levels described by
the projection operators
$\ket{g}\bra{g}$ representing the ground state and
$\ket{e}\bra{e}$ representing the excited state.
When the system is prepared at $t=0$ to be in the excited state,
$\ket{e}$, and it is coupled to a correctly tuned
radiation field,
the Born probability at $t$ to find the system in the ground state,
$\ket{g}$, is
\begin{equation}
\label{born_rabi}
\mathcal{P}_{\ket{g}\bra{g}}\big(\rho(t)\big)= \textrm{sin}^2(\Omega t)=
\frac{1}{2}\big(1-\textrm{cos}(2 \Omega t)\big),
\end{equation}
where $\Omega$ is called the Rabi frequency.
(The probability to find the system in the same state in which it was prepared
is $\textrm{cos}^2(\Omega t)$.)
These expressions are calculated for closed systems using standard quantum mechanics.
The $t$ in \eqref{born_rabi} is therefore
a single time parameter rather than a time coordinate or a set, $\{t\}$,
of time parameters.

In actual experiments, one never measures the result in~\eqref{born_rabi},
because in practice, a physical system cannot be isolated from its environment.
In this paper we wish to consider several very different
but very clean and well-controlled
experiments~\cite{meekhof_rabi_1996,brune_rabi_1996,petta_coherent_2005,
cole_nature_2001,zrenner_nature_2002,wang_macdonald_prb_2005,ramsay_prl_2010}.

In one experiment~\cite{meekhof_rabi_1996},
internal levels of a $^9$Be$^+$ ion couple to
a harmonic binding potential.
Rabi oscillations occur between two of the coupled internal and vibrational
levels, and the oscillations between different sets of levels are measured.
In another experiment~\cite{brune_rabi_1996},
Rabi oscillations are observed between the
circular states of a Rydberg atom coupled to a field stored in
a high $Q$ cavity.
In~\cite{petta_coherent_2005},
the Rabi oscillations are between the spin states of
two electrons in a double quantum dot.
In~\cite{cole_nature_2001}, Rabi oscillations occur between motional states
of electrons bound to shallow donors in semiconductors.
These motional states mimic the levels of a hydrogen atom.
In three experiments~\cite{zrenner_nature_2002,wang_macdonald_prb_2005,ramsay_prl_2010},
oscillations occur between excitation levels of electrons confined in quantum dots.

In~\cite{meekhof_rabi_1996,brune_rabi_1996,petta_coherent_2005},
one does not match the prediction in~\eqref{born_rabi}.
Instead, the measured probability is fit by an appropriately damped sinusoid,
\begin{equation}
\label{damped_rabi}
\mathcal{P}_{\ket{g}\bra{g}}\big(\rho(\tilde{t})\big)=
\frac{1}{2}\big(1-e^{-\gamma \tilde{t}}\,\textrm{cos}(2 \Omega \tilde{t})\big),
\end{equation}
where $\gamma$ is an experimentally determined damping factor.
Dynamical measurements are made in the lab, and
the $\tilde{t}$ in \eqref{damped_rabi} is therefore a time coordinate.
For simplicity, we have begun to set $\tilde{t}_{prep}=0$, with the understanding
that $\tilde{t}$ hereafter represents a duration from $\tilde{t}_{prep}$.

In~\cite{petta_coherent_2005}, amplitude, offset, and phase are fit as well,
because the oscillations are measured also as a function of a swept detuning voltage.
In~\cite{wang_macdonald_prb_2005,ramsay_prl_2010}, photocurrent is measured, and
from it the decoherence of the Rabi oscillations is inferred.
Numerically solving the models they fit to their data, the subsystems undergoing 
Rabi oscillations fit~\eqref{damped_rabi}.

Experiments also reveal that the damping factor, $\gamma$, depends
generally on the Rabi frequency, $\Omega$, regardless of the nature of the
experiment.
In the experiment with $^9\textrm{Be}^+$ ions~\cite{meekhof_rabi_1996},
the different levels are described by the kets
$\ket{\downarrow, n}$ and $\ket{\uparrow, n+1}$, where
$\ket{\downarrow}$ and $\ket{\uparrow}$ are internal states of the Be ion, and
$\ket{n}$ represents vibrational Fock states.
Rabi oscillations are measured for the
frequencies~\cite{wineland_experimental_1998,meekhof_rabi_1996}
\begin{equation}
\label{eq:freq_seq}
\Omega_{n,n+1} = \Omega \frac{0.202 \,e^{-0.202^2/2}}{\sqrt{n+1}} L^1_n(0.202^2),
\end{equation}
where $L^1_n$ is the generalized Laguerre polynomial.
The corresponding damping factor, $\gamma_n$, is measured
to increase with $n$ according to 
\begin{equation}
\label{eqn:gamma_meas}
\frac{\gamma_n}{\gamma_0}\approx (1+n)^{0.7}. \qquad \textrm{(Measured)}
\end{equation}

In the experiment~\cite{petta_coherent_2005}
with spin states of two electrons in a double quantum dot,
the damping factor, $\gamma$, is stated to be proportional to the
Rabi frequency, $\Omega$, though no mathematical relation is given.

In~\cite{cole_nature_2001,zrenner_nature_2002,wang_macdonald_prb_2005,ramsay_prl_2010},
this dependence has been called Excitation Induced Dephasing (EID), and its cause
has remained an open question.
In~\cite{wang_macdonald_prb_2005,ramsay_prl_2010}, the dependence of the
damping factor on the Rabi frequency has been found to be $\gamma\propto\Omega^2$.

These results for such different experiments
strongly suggest that some dependence of $\gamma$ on $\Omega$ is general
and independent of experimental specifics.

\subsection{Development of a Model}
Our task is to develop a single model of decoherence that reproduces
\eqref{damped_rabi} for systems undergoing Rabi oscillations, and, if
possible, the general dependence of $\gamma$ on $\Omega$, or EID,
also mentioned above.
Because of the distinction between parameters, $t$, and coordinates, $\tilde{t}$,
measured results made in coordinate time must be compared with
probabilities calculated also in coordinate time.
We will therefore call $\mathcal{P}_{\Lambda}\big(\rho(\tilde{t})\big)$,
which is a function of coordinate time, the ``predictive probability.''
And below we will take $\Lambda=\ket{g}\bra{g}$.

Using the branched density operator from~\eqref{eq:branched}, we have
the predictive probability
\begin{eqnarray}
\label{eq:long_prob}
\mathcal{P}_{\Lambda}\big(\rho(\tilde{t})\big) & = &
\Tr\big( \Lambda \rho(\tilde{t})\big)\Leftrightarrow
\Tr\big( \Lambda \rho(\{t\})\big) \nonumber \\
& = & \Tr \big( \Lambda \,a_1\, \rho_\phi(t_1) +
\Lambda \,a_2\, \rho_\chi(t_2) + \ldots\big)  \nonumber \\
& = & a_1\,\mathcal{P}_\Lambda\big(\rho_\phi(t_1)\big) +
      a_2\,\mathcal{P}_\Lambda\big(\rho_\chi(t_2)\big) + \ldots \,\,\,\,\,\,
\end{eqnarray}

The long-time behavior of the measured result in~\eqref{damped_rabi} is
$\mathcal{P}_{\ket{g}\bra{g}}\big(\rho(\tilde{t})\big)=
\mathcal{P}_{\ket{e}\bra{e}}\big(\rho(\tilde{t})\big)\rightarrow\frac12$.
From this we can deduce that there is no dissipation for the experimental system,
and that the term $\mathcal{P}_\Lambda\big(\rho_\chi(t_2)\big)$
in~\eqref{eq:long_prob} ought not to introduce losses.
For these clean experiments, we then hypothesize that 
immediately after the uncontrolled occurrence, $\Phi_\phi=\Phi_\chi$.
Therefore, we want a predictive probability of the form
\begin{equation}
\label{eq:sum_branch}
\mathcal{P}_{\Lambda}\big(\rho(\tilde{t})\big) \Leftrightarrow \sum_{m=0}^{M_{branch}}\,
      a_m\,\mathcal{P}_\Lambda\big(\rho_m(t_m)\big)
\end{equation}
where $M_{branch}$ is the number of branches.
Note that we no longer require the label $\phi$ for $\rho$.
This formula is deceptively simple in appearance, because, in general,
the number of branches, $M_{branch}$,
will depend on $\tilde{t}$, and the value of $a_{m^\prime}$ 
for a fixed $m^\prime$ will depend on $M_{branch}$.

The time parameter values $t_m$ are related unambiguously to $\tilde{t}$ 
via the rule in Section~\ref{sec:ensembles}.
The $a_m$ will come from models.

We still require a prescription for the $\rho_m$.
Leggett has explained~\cite{leggett_arrow_qmeasurement} that decoherence
``is exactly the result of a `measurement' whose result is uninspected,''
and he has called this process ``garbling.''
Measuring a system to be in a state
is equivalent to preparing that system to be in that state, so
these passive measurements that garble an ensemble are also passive preparations.
We will therefore assume that the effect of an uncontrolled interference event
is to measure and therefore \emph{prepare anew} some subset of our ensemble.
And for obvious reasons, one should assume that nature does not distinguish between
passive and active preparations.

A hypothesis physically equivalent to garbling is that the uncontrolled environment
interacts only with energy eigenstates of the system of interest.
Then, because $\Phi_\chi=\Phi_\phi$, when the uncontrolled event occurs
at $t_m=0$, the most general form for $\rho_m$ is
\begin{equation}
\label{eq:garb1}
\rho_m(t_m=0)=w_m\,\,\ket{g}\bra{g} + v_m\,\, \ket{e}\bra{e},
\end{equation}
with $w_m+v_m=1$.
In~\eqref{eq:garb1}, $w$ is a weight appropriate for ensemble members 
prepared passively to be in the ground state, and $v$ is a weight for 
ensemble members prepared in the excited state.
The weights are of course independent of $t_m$, and
their values will also come from our model.

\subsection{Predictive Probability}
\label{sec:dist_dev}
Given our hypotheses and deductions, the simplest
reasonable and general model the author can conceive is:
\begin{enumerate}
\item Members of the experimental ensemble
are actively prepared in the lab at $\tilde{t}=\tilde{t}_{prep}$.
Again, for simplicity, we set $\tilde{t}_{prep}=0$.
\item The members of the ensemble can possibly
suffer environmental interactions at the times
$n\Delta\tilde{t}$, where $n=1,2,3\ldots$
As a result, some systems are passively prepared to be in either the ground
state or the excited state.
We will call these interactions ``interference events.''
\item At every $n\Delta\tilde{t}$, there is some probability, $(1-\eta$),
for a member to suffer perturbation and thereby to be prepared passively.
\end{enumerate}
In the second step, we have introduced the time scale, $\Delta\tilde{t}$,
of interaction with the environment.

In the third step we have introduced a parameter, $\eta$, with $0\leq\eta\leq 1$,
to represent the susceptibility of ensemble members to environmental interference.
As described, the parameter $\eta$ is the probability for any given
member of the ensemble \emph{not to suffer} interference at one of the
times $n\Delta \tilde{t}$.
For a perfectly isolated system, $\eta=1$.

According to this simple model,
we can write a general formula for the predictive probability
as more and more branches form:
\begin{equation}
\label{prob_written}
\mathcal{P}_{\ket{g}\bra{g}}\big(\rho(\tilde{t})\big) =
\begin{cases}
p_0(\tilde{t}) & 0\leq\tilde{t}< 1\,\Delta\tilde{t} \\
p_1(\tilde{t}) & 1\,\Delta\tilde{t}\leq\tilde{t}< 2\,\Delta\tilde{t} \\
\quad\vdots & \qquad\vdots \\
p_n(\tilde{t}) & n\Delta\tilde{t}\leq\tilde{t}< (n+1)\Delta\tilde{t} 
\end{cases}
\end{equation}
Here, every $p_n(\tilde{t})$ will have
the form of the sum in~\eqref{eq:sum_branch}, and $M_{branch}=n$.

Let us assume that a physicist identically prepares every member of the
experimental ensemble initially to be in the excited state, $\ket{e}$.
Because no members will have suffered environmental interference before
$\tilde{t}=1\,\Delta\tilde{t}$, we have for the initial value
$p_0(\tilde{t})=\textrm{sin}^2(\Omega \tilde{t})$.

At $\tilde{t}=1\,\Delta\tilde{t}$, the fraction $(1-\eta)$
of the ensemble will be passively measured and thus prepared 
in a new state, with new initial conditions.
At this point, $M_{branch}=1$, and from~\eqref{eq:sum_branch} we have
\begin{equation}
\label{eq:b1}
p_1(\tilde{t})\Leftrightarrow
a_0\,\mathcal{P}_{\ket{g}\bra{g}}\big(\rho_0(t_0)\big) + 
a_1\,\mathcal{P}_{\ket{g}\bra{g}}\big(\rho_1(t_1)\big).
\end{equation}
The first term in~\eqref{eq:b1},
$a_0\,\mathcal{P}_{\ket{g}\bra{g}}\big(\rho_0(t_0)\big)$,
corresponds to the subset of the ensemble that passes the time $1\Delta\tilde{t}$
without suffering interaction with the environment.
According to our model, it is 
$\eta\,\textrm{sin}^2(\Omega\tilde{t})=\eta\,p_0(\tilde{t})$.
For this \emph{unperturbed} subset, of course, $\tilde{t}\Leftrightarrow t_0$.

The second term in~\eqref{eq:b1},
$a_1\,\mathcal{P}_{\ket{g}\bra{g}}\big(\rho_1(t_1)\big)$, is the first branch.
It corresponds to the subset of the ensemble that suffers perturbation and is
thereby garbled, or passively prepared anew.
According to our model, $a_1=(1-\eta)$.

By our prescription for the density operators, we know that
$\rho_1(t_1=0)$ is fixed by the state
of the perturbed subset as it is prepared at $\tilde{t}=1\Delta\tilde{t}$
by the uncontrolled event.
The perturbed subset is passively measured at $\tilde{t}=1\Delta\tilde{t}$, so
assuming that garbling is a fair measurement,
$\rho_1(t_1=0)$ will have the form of~\eqref{eq:garb1} with
$w_1=\textrm{sin}^2(1\,\Omega \Delta \tilde{t})=p_0(1\Delta\tilde{t})$ and
$v_1=1-p_0(1\Delta\tilde{t})$.

Finally, with the identification
$t_1\Leftrightarrow \tilde{t}-1\Delta\tilde{t}$ for the perturbed subset
(see Figure~\ref{fig:timeline}), it is straightforward to find
from the theory of Rabi oscillations
\begin{widetext}
\begin{equation}
\label{eq:first_rec}
p_1(\tilde{t})=\eta\,p_0(\tilde{t})+
(1-\eta)
\Big(p_0(1\Delta\tilde{t})\,\textrm{cos}^2\big(\Omega\,(\tilde{t}-1\Delta\tilde{t})\big)
+ \big(1-p_0(1\Delta\tilde{t})\big)\,\textrm{sin}^2\big(\Omega\,(\tilde{t}-1\Delta\tilde{t}) \big)
 \Big).
\end{equation}
\end{widetext}

This is a nice form because we can stop writing out branches and use a recursive
formula for the predictive probability.
For general $n$, we have
\begin{widetext}
\begin{equation}
\label{eq:general_rec}
p_n(\tilde{t})=\eta\,p_{n-1}(\tilde{t})+
(1-\eta)
\Big( p_{n-1}(n\Delta\tilde{t})\,\textrm{cos}^2\big(\Omega\,(\tilde{t}-n\Delta\tilde{t})\big)
+\big(1-p_{n-1}(n\Delta\tilde{t})\big)\,\textrm{sin}^2\big(\Omega\,(\tilde{t}-n\Delta\tilde{t})
\big) \Big).
\end{equation}
\end{widetext}

\begin{figure}[htp]
  \begin{center}
    \subfigure[With $\eta=0.99$, we fit $\gamma/\Omega=0.05$.]
    {\label{rec1}\includegraphics[width=0.45\textwidth]{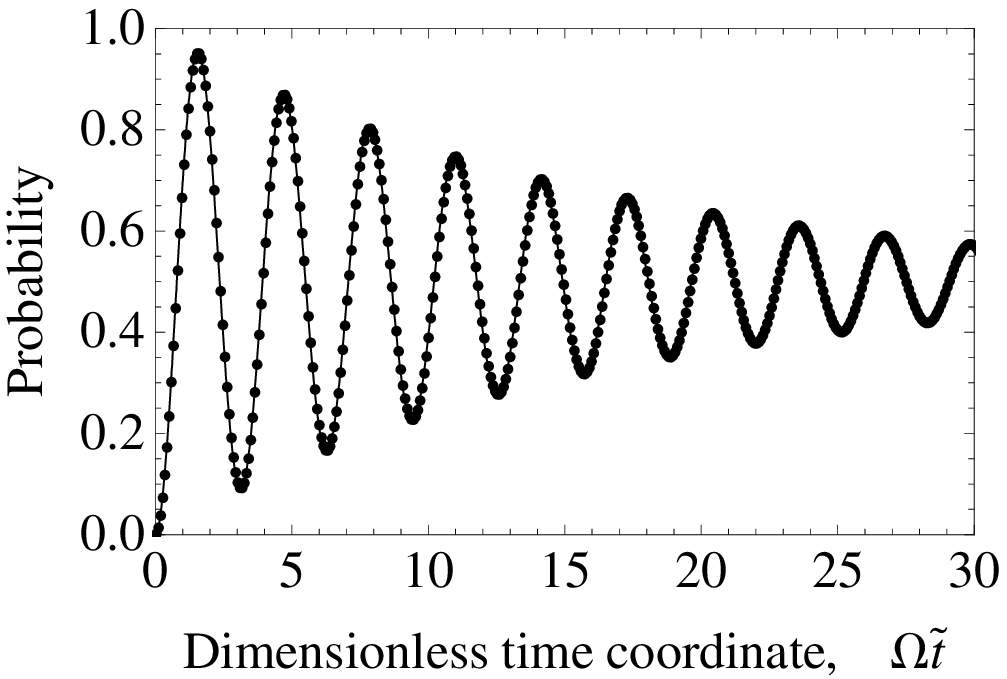}}\hfill
    \subfigure[With $\eta=0.997$, we fit $\gamma/\Omega=0.015$.]
    {\label{rec2}\includegraphics[width=0.45\textwidth]{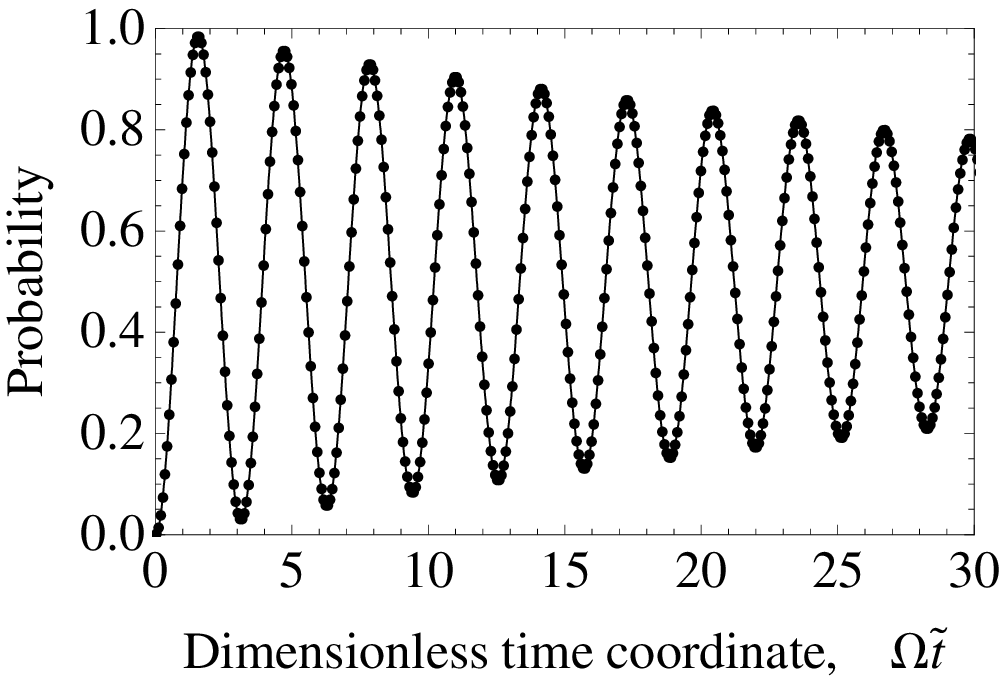}}\\
  \end{center}
  \caption{The predictive probability,
           $\mathcal{P}_{\ket{g}\bra{g}}\big(\rho(\tilde{t})\big)$,
           when members of the ensemble are distinguishable.  For both plots,
           $\Omega\Delta\tilde{t}\approx 0.08$.
           The dots are the results of recursive calculations
           using \eqref{eq:general_rec} and \eqref{prob_written}.
           The solid lines are plots of the damped sinusoid
           \eqref{damped_rabi} that fits the experimental data.
           Recall that $\eta$ is the probability that members
           will not suffer a perturbation at the times $n\Delta\tilde{t}$.
           Values of $\eta$ and $\Delta\tilde{t}$ were chosen for
           aesthetics only.
           These plots do look good, but our results are not yet correct.
           See the text.}
  \label{fig:recursive}
\end{figure}

Figure~\ref{fig:recursive} shows the results of two sample calculations using
\eqref{prob_written} with \eqref{eq:general_rec}.
The dots are from the recursive calculation using our model.
The solid lines are plots of the decaying sinusoid in \eqref{damped_rabi},
which fits the experimental measurements.
For both figures we have used $\Omega\Delta\tilde{t}\approx 0.08$.
The results in Figure~\ref{rec1} were calculated using $\eta=0.99$
and resulted in a fitted value for the damping factor of
$\gamma/\Omega=0.05$.
The results in Figure~\ref{rec2} were calculated using $\eta=0.997$
and resulted in a fitted value for the damping factor of
$\gamma/\Omega=0.015$.
Recall that $\eta=1$ for a perfectly isolated system.

Our calculation of the predictive probability results in no frequency shift
away from the Rabi frequency, $\Omega$, at early times,
in agreement with experiments.
(With the master equation, one predicts a frequency shift
and must assume very strong driving to get rid of it.
See Section~\ref{sec:compare}.)
Note, however, that our method results in a fitted
damping factor, $\gamma$,
that is independent of $\Omega$.
Though our first attempt has been instructive, it is thus not correct.
We have included this incorrect result not only because it clearly
illustrates our new method, but also because
the recursive structure of \eqref{eq:general_rec} will
reappear in the correct model below.

\subsection{Predictive Probability and Indistinguishability}
\label{sec:prob_indist}
Consider the experiments~\cite{meekhof_rabi_1996,brune_rabi_1996,petta_coherent_2005}
in which one repeatedly prepares and then measures single members of the experimental
ensemble.
One assumes that performing a number of consecutive
measurements, each done on a single member,
is equivalent to performing a measurement on the same number
of non-interacting, simultaneously present members of an identically prepared
ensemble.
This is a statement of ergodicity.
And when the identically prepared members are present simultaneously,
one treats them as indistinguishable.
In our first model, we did not treat them so.
And to enforce ergodicity, we would like to.

Though in these experiments, one can easily label any member of the ensemble with 
the value of the time coordinate, $\tilde{t}$, at which it was
actively prepared or measured,
coordinate values are always physically meaningless.
The only physically significant label for identically prepared
members is the \emph{duration} between preparation and measurement.
And, even in these experiments, one \emph{cannot in principle} label any given
member of an ensemble with a duration.
To understand why, consider the result of a hypothetical, active
measurement on any single member of an experimental ensemble:
\begin{enumerate}
\item At $\tilde{t}=0$ a physicist actively prepares the members.
\item At $\tilde{t}=1\Delta\tilde{t}$ subset $A$ is passively prepared
to be in a mixture of the ground state and the excited state.
\item At $\tilde{t}=2\Delta\tilde{t}$ subset $B$ is passively prepared
to be in a mixture of the ground state and the excited state.
\item At some time coordinate value $\tilde{t} > 2\Delta\tilde{t}$
a physicist actively measures a single member
to be in the ground state, $\ket{g}$.
\end{enumerate}
Because of the probabilistic nature of quantum mechanics, it is
impossible to distinguish if the actively
measured member was in subset $A$, subset $B$, both, or neither.

In other words, if the state of an ensemble is represented by
the branched density operator
\begin{equation}
\label{eq:branched_demo}
\rho(\{t\}) =
a_0\,\,\rho_0(t_0) + a_A\,\,\rho_A(t_A) + a_B\,\,\rho_B(t_B),
\end{equation}
then it is impossible to know to which branch any given member belongs.

One way to distribute indistinguishable ensemble members among several
branches is to count different combinations of the members themselves.
But we cannot use such an approach for two important reasons.
The first is because quantum mechanics makes predictions for ensemble
averages, and density operators simply do not follow the trajectories
of individual members.
The second reason is that any model based on such counting will depend
on the size of the experimental ensemble.
This is clearly unacceptable.

If, however, one imagines that ensemble members are labeled by the
individual interference events at which they have been most recently 
prepared, we hypothesize that we can enforce ergodicity by treating 
the events themselves as indistinguishable.
The resulting models will depend not on the ensemble size but instead
on the number of interference events, which is exactly what we want.

When modeling experiments, one would probably use
a full simulation.
To maintain transparency,
however, we will again search for an analytical formula.
First, assume that we wish to calculate the predictive probability
given that there have been $n$ chances for ensemble members to have
suffered a single interference event.
If interference events can possibly occur at intervals of
$\Delta\tilde{t}$, the predictive probability can be written
$\mathcal{P}_{\ket{g}\bra{g}}\big(\rho(n\Delta\tilde{t})\big)$.
Our scenario is as follows:
\begin{enumerate}
\item\label{choice1} Choose a combination of time intervals
(of length $\Delta\tilde{t}$) such that systems have
survived only a duration of $1\,\Delta\tilde{t}$ before being
passively measured.
\item\label{adj1} Adjust the probability \textbf{at}
$\boldsymbol n \boldsymbol\Delta \tilde{\boldsymbol t}$ accordingly.
\item Because the systems and interference events are indistinguishable,
so are the intervals between the events.
Put the intervals chosen above back into the original set.
\item\label{choice2} Choose a combination of time intervals
such that systems have
survived a duration of $2\,\Delta\tilde{t}$ before being
passively measured.
\item\label{adj2} Adjust the probability \textbf{at}
$\boldsymbol n \boldsymbol\Delta \tilde{\boldsymbol t}$ accordingly.
\item Put the intervals chosen above back into the original set.
\item Repeat until reaching $n\Delta\tilde{t}$.
\end{enumerate}
We have again chosen to parametrize our model with a time
scale, $\Delta\tilde{t}$.
However, we will require a slightly different interpretation
for the number parameterizing the members' susceptibility to environmental interference.
We shall now use $\beta$ (rather than $\eta$), with $0\le \beta\leq 1$ and
$\beta=1$ for a perfectly isolated system.
Physically, $\beta$ will be the probability that a randomly chosen time interval
will have come before an interference event.
This can be better understood after equation \eqref{eq:mom}.

To implement steps \ref{choice1} and \ref{choice2} above,
we will make use of the binomial distribution,
\begin{equation}
\label{eq:bin_dist}
b(n,k,\beta) \equiv
{n \choose k} \beta^k (1-\beta)^{n-k}.
\end{equation}
The distribution in \eqref{eq:bin_dist} gives
the probability for the occurrence of any combination, regardless
of order, of $k$ events with probability $\beta$ and $(n-k)$ events
with probability $(1-\beta)$.
Because the distribution gives probabilities for combinations rather
than permutations, with it we can treat indistinguishable interference
events.
Note also the normalization
\begin{equation}
\label{eq:binorm}
\sum_{k=0}^{n} {n \choose k} \beta^k (1-\beta)^{n-k} = 1.
\end{equation}
Using \eqref{eq:binorm} we will relate the binomial distribution to a probability.

To implement steps \ref{adj1} and \ref{adj2} in this model,
we will once more use the rule
deduced in Section~\ref{sec:ensembles}, to reset to $t=0$
the time evolution parameter of the density operator corresponding to members
passively prepared as a result of environmental interference.

To see how the binomial weights work, let us write a simple formula for
the case that individual ensemble members will have suffered \emph{only one}
interference event before an active measurement occurs.
This will truncate our formula at a reasonable size, and we will generalize
to multiple events below.
The probability at $4\Delta\tilde{t}$
to find in $\ket{g}$ members that have been initially
prepared at  $\tilde{t}=0$ to be in $\ket{e}$ is
\begin{widetext}
\begin{eqnarray}
\label{eq_simple_case}
\mathcal{P}_{\ket{g}\bra{g}}\big(\rho(4\Delta\tilde{t})\big) & = &
    b(4,4,\beta)\Big(\textrm{sin}^2(\Omega\, 4\Delta\tilde{t})\,\textrm{cos}^2(\Omega\, 0\Delta\tilde{t})
 + \textrm{cos}^2(\Omega\, 4\Delta\tilde{t})\,\textrm{sin}^2(\Omega\, 0\Delta\tilde{t})\Big)+\nonumber \\
& & b(4,3,\beta)\Big(\textrm{sin}^2(\Omega \,3\Delta\tilde{t})\,\textrm{cos}^2(\Omega \,1\Delta\tilde{t})
 + \textrm{cos}^2(\Omega \,3\Delta\tilde{t})\,\textrm{sin}^2(\Omega \,1\Delta\tilde{t})\Big)+\nonumber \\
& & b(4,2,\beta)\Big(\textrm{sin}^2(\Omega \,2\Delta\tilde{t})\,\textrm{cos}^2(\Omega \,2\Delta\tilde{t})
 + \textrm{cos}^2(\Omega \,2\Delta\tilde{t})\,\textrm{sin}^2(\Omega \,2\Delta\tilde{t})\Big)+\nonumber \\
& & b(4,1,\beta)\Big(\textrm{sin}^2(\Omega \,1\Delta\tilde{t})\,\textrm{cos}^2(\Omega \,3\Delta\tilde{t})
 + \textrm{cos}^2(\Omega \,1\Delta\tilde{t})\,\textrm{sin}^2(\Omega \,3\Delta\tilde{t})\Big)+\nonumber \\
& & b(4,0,\beta)\Big(\textrm{sin}^2(\Omega \,0\Delta\tilde{t})\,\textrm{cos}^2(\Omega \,4\Delta\tilde{t})
 + \textrm{cos}^2(\Omega \,0\Delta\tilde{t})\,\textrm{sin}^2(\Omega \,4\Delta\tilde{t})\Big).
\end{eqnarray}
\end{widetext}

The explanation of \eqref{eq_simple_case} is straightforward.
We need to relate the binomial distribution to the passage of time, so
at every step $k$, with $0\leq k\leq n=4$, we will count the 
(normalized) number of combinations for arranging the $n$
time intervals such that $k$ of them came before the single interference event.
This number is given by $b(n,k,\beta)$.
Because possible interference events occur at increments of the time
scale, $\Delta\tilde{t}$, the weight $b(n,k,\beta)$ must then be attached
to any passive preparation occurring at $k\Delta\tilde{t}$.

Again, branches of a density operator form when subsets of the ensemble
suffer environmental interference and are passively prepared.
For a branched density operator, in each line of~\eqref{eq_simple_case},
the term multiplied by $b(n,k,\beta)$ is the
Born probability for the branches formed at $k\Delta\tilde{t}$.

Consider the second line, where $k=3$, and compare
to the notation in~\eqref{eq:garb1}.
The weight $w=\textrm{sin}^2(\Omega \,3\Delta\tilde{t})$.
The weight $v=\textrm{cos}^2(\Omega \,3\Delta\tilde{t})$.
The \emph{duration} from passive preparation to $4\Delta\tilde{t}$ is then
$(4-3)\Delta\tilde{t}=1\Delta\tilde{t}$, so the functions with argument
$1\Delta\tilde{t}$ contain the time dependence of the Born probability
of the mixed state.


Let us introduce the notation
$\mathcal{P}^{(i)}_{\ket{g}\bra{g}}\big(\rho(n\Delta\tilde{t})\big)$
to represent the predictive probability under the assumption
that members on average will have suffered 
$i$ or fewer interference events before measurement.
Then for general $n$,
\begin{widetext}
\begin{equation}
\label{eq:one_event}
\mathcal{P}^{(1)}_{\ket{g}\bra{g}}\big(\rho(n\Delta\tilde{t})\big) =
\sum_{k=0}^n b(n,k,\beta)
\Big(\textrm{sin}^2(\Omega\,k \Delta\tilde{t})\,\textrm{cos}^2(\Omega(n-k)\Delta\tilde{t})
 + \textrm{cos}^2(\Omega\,k\Delta\tilde{t})\,\textrm{sin}^2(\Omega(n-k)\Delta\tilde{t})\Big).
\end{equation}
\end{widetext}
By simply exchanging the
$\textrm{cos}^2(\Omega\, k\Delta\tilde{t})$ and
$\textrm{sin}^2(\Omega\, k\Delta\tilde{t})$ terms,
we calculate
$\mathcal{P}^{(1)}_{\ket{e}\bra{e}}\big(\rho(n\Delta\tilde{t})\big)$,
which is the probability to find the ensemble members
in the excited state.

In \eqref{eq:one_event},
we have assumed that systems will have suffered at most one interference event.
To allow for the possibility of multiple events, the terms
$\textrm{sin}^2(\Omega\, k\Delta\tilde{t})$ and
$\textrm{cos}^2(\Omega\, k\Delta\tilde{t})$ 
must be replaced with new functions of $k\Delta\tilde{t}$,
that predict the effects of interference events prior to the
single event assumed in \eqref{eq:one_event}.
For general $i$,
\begin{widetext}
\begin{equation}
\label{eq:two_events}
\mathcal{P}^{(i)}_{\ket{g}\bra{g}}\big(\rho(n\Delta\tilde{t})\big) = \sum_{k=0}^n b(n,k,\beta)
\Big(\mathcal{P}^{(i-1)}_{\ket{g}\bra{g}}\big(\rho(k\Delta\tilde{t})\big)\,
\textrm{cos}^2(\Omega(n-k)\Delta\tilde{t})
 + \mathcal{P}^{(i-1)}_{\ket{e}\bra{e}}\big(\rho(k\Delta\tilde{t})\big)\,
\textrm{sin}^2(\Omega(n-k)\Delta\tilde{t})
\Big).
\end{equation}
\end{widetext}

In \eqref{eq:two_events}, the recursive structure of
\eqref{prob_written} and \eqref{eq:general_rec} has reappeared,
but here the ensemble members and interference events are indistinguishable.
Handling the possibility for more and more interference events requires
the nesting of more and more summed terms into \eqref{eq:two_events}.
These nested equations are significantly more difficult to solve
than are the recursive equations used for distinguishable events.

We have written the predictive probability as a function of $n\Delta\tilde{t}$.
The final step is to scale our result back to the time coordinate, $\tilde{t}$.
The first moment of the binomial distribution is
\begin{equation}
\label{eq:mom}
\langle k \rangle = \sum_{k=0}^{n} {n \choose k} \beta^k (1-\beta)^{n-k} \,k =\beta\, n.
\end{equation}
After stepping through time to $n\Delta\tilde{t}$,
on average $\beta\,n$ of the intervals will have preceded
the interference event number $i$.
This provides us a physical interpretation of our two parameters,
and to ensure that the time scales with something physical, we will need to use
$\langle k \rangle \Delta\tilde{t}= \beta\, n\Delta\tilde{t}=\tilde{t}$.
After a calculation of the predictive probability as a function of
$n\Delta\tilde{t}$, we make the replacement
\begin{equation}
n\rightarrow \frac{\tilde{t}}{\beta \,\Delta\tilde{t}}\,.
\end{equation}
This restricts us to non-zero values of $\beta$ and $\Delta\tilde{t}$.
We have also simply interpolated between the discrete
values of $n$ at which \eqref{eq:two_events} is actually defined.
Because our time scale is understood to be an average value,
it would be inappropriate to assume for our model anything more complicated.

\begin{figure}[h]
\includegraphics[width=.5\textwidth]{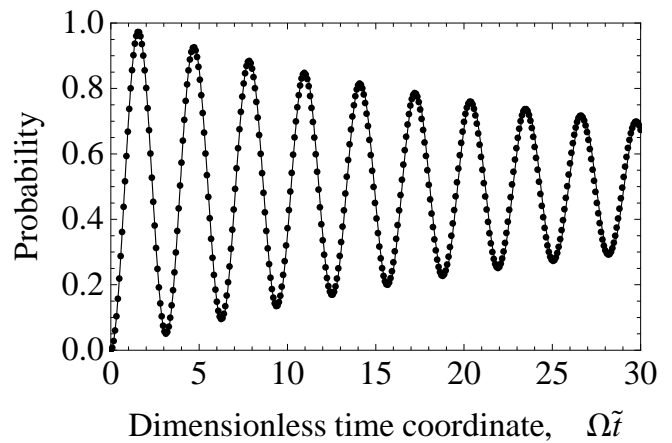}
\caption{Plot of the predictive probability,
$\mathcal{P}^{(5)}_{\ket{g}\bra{g}}\big(\rho(\tilde{t})\big)$,
for indistinguishable ensemble members.
For the dots we have used equations \eqref{eq:two_events}
and \eqref{eq:subst}.
The solid line is a plot of the damped sinusoid \eqref{damped_rabi}
that fits the experimental data.
We have used $\Omega\Delta\tilde{t}\approx 0.7$ and $\beta=0.995$,
and we have fit $\gamma / \Omega = 0.039$.}
\label{fig:indis}
\end{figure}

The predictive probability at $\tilde{t}$,
assuming $i$ interference events, to find 
in the ground state a system
initially prepared in the excited state is therefore
\begin{equation}
\label{eq:subst}
\mathcal{P}^{(i)}_{\ket{g}\bra{g}}\big(\rho(\tilde{t})\big)=
\mathcal{P}^{(i)}_{\ket{g}\bra{g}}\big(\rho(n\Delta\tilde{t})\big),\quad
n\rightarrow\frac{\tilde{t}}{\beta\,\Delta\tilde{t}}\,.
\end{equation}
Numerical solution of \eqref{eq:two_events} is straightforward.
One can also perform the summations
and find closed form expressions for the predictive probabilities.
In Figure \ref{fig:indis} we have plotted 
$\mathcal{P}^{(5)}_{\ket{g}\bra{g}}\big(\rho(\tilde{t})\big)$,
which is the predictive probability given that there have been for the average
member $5$ or fewer interference events.
For clean experiments, with $\beta$ close to $1$,
we expect a good approximation
for $i=5$.
Again, the agreement with the experimentally measured damped sinusoid is very good.

As opposed to the previous results for distinguishable ensemble members, however,
when we treat the interference events as indistinguishable, we
find that the damping factor, $\gamma$, \emph{does in general} depend on the 
Rabi frequency, $\Omega$, in agreement with measurements and the phenomenon
called Excitation Induced Dephasing.
Furthermore, in our simple model we have assumed
only a time scale, $\Delta\tilde{t}$, and a parameter, $\beta$, 
describing the experimental system's susceptibility to interference from
the environment.
It can be applied to \emph{any} real system undergoing Rabi oscillations.

For $\beta$ close to $1$, the predictive probability is dominated by the
terms proportional to $b(n,k=n,\beta)$.
Solving the rather crude, truncated form
\begin{equation}
\mathcal{P}_{\ket{g}\bra{g}}\big(\rho(n\Delta\tilde{t})\big) \approx \sum_{k=0}^n
b(n,k,\beta)\textrm{sin}^2(\Omega\,k\Delta\tilde{t})
\end{equation}
and using \eqref{eq:subst}, we get
\begin{eqnarray}
\mathcal{P}_{\ket{g}\bra{g}}\big(\rho(\tilde{t})\big)& \approx & 
\frac{1}{4} \Big(2-
\big( 1-\beta (1-e^{-2 i \Delta\tilde{t}\Omega})\big)^\frac{\tilde{t}}{\beta\Delta\tilde{t}}
\nonumber \\
\label{eq:approx}
& &\,\,\quad-\big( 1-\beta
(1-e^{+2 i \Delta\tilde{t}\Omega})\big)^\frac{\tilde{t}}{\beta\Delta\tilde{t}}
\Big).
\end{eqnarray}
For small $\Delta\tilde{t}$, \eqref{eq:approx} is
\begin{equation}
\mathcal{P}_{\ket{g}\bra{g}}\big(\rho(\tilde{t})\big)=
\frac{1}{2}\Big(1-e^{-\gamma \tilde{t}}\,\big(\textrm{cos}(2 \Omega \tilde{t})
+\textrm{O}(\Delta\tilde{t}^{\,2})\big)\Big),
\end{equation}
where
\begin{equation}
\gamma = 2\, (1-\beta)\, \Omega^2\, \Delta\tilde{t} + \textrm{O}(\Delta\tilde{t}^{\,3}).
\end{equation}
For $\beta \approx 1$ and $\Omega\Delta\tilde{t} \ll 1$, 
$\gamma$ is quadratic in $\Omega$.
This matches the results in~\cite{wang_macdonald_prb_2005,ramsay_prl_2010}.
And when the systems are treated as indistinguishable, we require none of
the typical assumptions regarding the frequency response of perturbation
terms~\cite{romito_decoherence_2007,wang_macdonald_prb_2005,ramsay_prl_2010,
mogilevtsev_prl_2008}.

\begin{figure}[h]
\includegraphics[width=.5\textwidth]{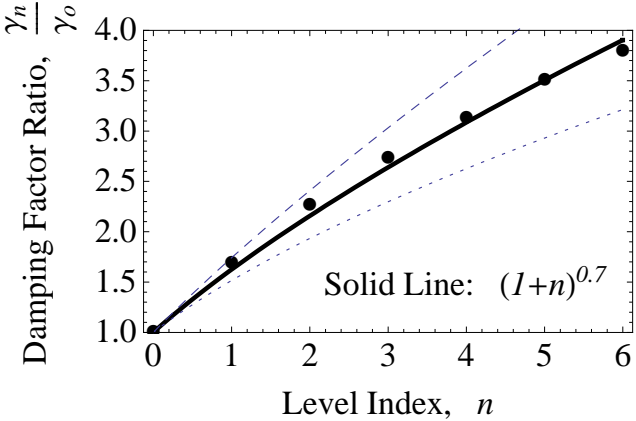}
\caption{Matching the experimental results for the ratio of
damping factors, $\frac{\gamma_n}{\gamma_0}$.
The large dots result from our theoretical calculation of the predictive
probability using \eqref{eq:two_events} and \eqref{eq:subst}.
The solid line is the experimentally measured relation, $(1+n)^{0.7}$.
To indicate a scale for the exponent,
the thin dashed line is a plot of $(1+n)^{0.8}$,
and the thin dotted line is a plot of $(1+n)^{0.6}$.}
\label{fig:fit}
\end{figure}

The same model is general enough to apply also 
to the system in~\cite{meekhof_rabi_1996},
and with it we fit the measured relation \eqref{eqn:gamma_meas}
between $\gamma$ and $\Omega$.
Figure~\ref{fig:fit} is the result of fitting the damping factor, $\gamma_n$, to 
$\mathcal{P}^{(5)}_{\ket{g}\bra{g}}\big(\rho(\tilde{t})\big)$,
calculated with the sequence of frequencies in \eqref{eq:freq_seq}
and with $\Omega_0\Delta\tilde{t}\approx 0.2$.
(The exponent can be shifted by choosing different time scales.)
Unfortunately, we are limited by computational resources to $n\leq 6$.

Figure~\ref{fig:fit} shows a very good quantitative agreement with
experiment \eqref{eqn:gamma_meas},
and again we have not required any experimentally specific assumptions.
Our study indicates that the decoherence measured in~\cite{meekhof_rabi_1996}
results from environmental interference having
a characteristic frequency of $\approx 5\Omega_0$.

Finally, we can conclude that a dependence of $\gamma$ on $\Omega$,
or EID, is indeed general, as suggested by experiment, and that it 
is a measurable effect
of the indistinguishability of separate, uncontrolled interactions
between quantum systems and their environment.

\subsection{Comparison With the Standard Program}
\label{sec:compare}
In the standard decoherence program, one treats these experiments
using the decoherence master equation.
It is the method of choice when one works
without time asymmetric boundary conditions and when one 
identifies $t$ with coordinate time.
The generic solution of the master equation describing a system
undergoing Rabi oscillations and on resonance
is~\cite{petruccione_open_quantum_systems}
\begin{equation}
\label{eq:mast_sol}
\mathcal{P}^{ME}_{\ket{g}\bra{g}}(t)=\frac{4\Omega^2}{\Gamma^2+8\Omega^2}
\big(1-e^{-3 \Gamma t/4}(\textrm{cos}\, \mu t + \frac{3\Gamma}{4 \mu}
\textrm{sin}\,\mu t)  \big).
\end{equation}
To match our convention, we have added a factor of $2$ to the
definition of Rabi Frequency in~\cite{petruccione_open_quantum_systems}.
In \eqref{eq:mast_sol}, $\mu=\sqrt{4\Omega^2-\big( \frac{\Gamma}{4} \big)^2}$.
Unless a specific environmental interaction has been assumed,
$\Gamma$ is the spontaneous emission rate
for the system in its excited state.
It is a constant, independent of the Rabi frequency.

A significant shift from the Rabi frequency, $\Omega$, is not
observed in experiments, however, so
one must assume very strong driving: $2\Omega\gg \Gamma/4$.
The limit of \eqref{eq:mast_sol} is then
\begin{equation}
\label{eq:mast_lim}
\mathcal{P}^{ME}_{\ket{g}\bra{g}}(t)= \frac{1}{2} \big(
1-e^{-3 \Gamma t/4} \textrm{cos}\,2\Omega t \big).
\end{equation}

One can see immediately why the measured results have been puzzling.
For general calculations,
the solution \eqref{eq:mast_lim} of the master equation predicts 
for the experiment in~\cite{meekhof_rabi_1996}
\begin{equation}
\label{eqn:gamma_meq}
\frac{\gamma_n}{\gamma_0}=1, \qquad\quad \textrm{(Master equation)}
\end{equation}
and that $\gamma$ is independent of $\Omega$.

In the hope of matching experiment,
one typically assumes a detailed form for the interaction operators.
For the experiment in~\cite{meekhof_rabi_1996}, in which~\eqref{eqn:gamma_meas}
was measured,
several such studies have been carried out~\cite{schneider_decoherence_1998,
murao_decoherence_1998,bonifacio_pra_2000,difidio_damped_2000,
serra_decoherence_2001,budini_localization_2002,budini_dissipation_2003}.
None of these studies, however, has resulted in an agreement as quantitatively good as
that in Figure~\ref{fig:fit}.

Furthermore, the models for decoherence in~\cite{schneider_decoherence_1998,
murao_decoherence_1998,bonifacio_pra_2000,difidio_damped_2000,
serra_decoherence_2001,budini_localization_2002,budini_dissipation_2003}
have necessarily been highly tuned.
They are thus not applicable to the other types of
experiments~\cite{petta_coherent_2005,cole_nature_2001,zrenner_nature_2002,
wang_macdonald_prb_2005,ramsay_prl_2010}. 
And, with the exception perhaps of~\cite{bonifacio_pra_2000}, they
suggest that a dependence
of $\gamma$ on $\Omega$ is not general, in disagreement with measurements.
Similarly, the 
models~\cite{romito_decoherence_2007,wang_macdonald_prb_2005,ramsay_prl_2010,
mogilevtsev_prl_2008}
used to explain the EID measured in
experiments with shallow donors and quantum dots cannot be applied to
ions in a Paul trap~\cite{meekhof_rabi_1996}.

This paper describes the general framework that one must use when applying a
time asymmetric theory.
Full treatment of experiments will likely require more detailed models
addressing multiple sources of environmental interference, dressed states, etc.
But the qualitative and quantitative success
of our single, general model for different types of Rabi oscillations experiments
is very promising.

\section{Conclusion}
A choice of boundary conditions results in intrinsic time asymmetry 
endowed at the microphysical level, even for closed systems.
The theoretical expression of this asymmetry is time evolution
generated by a semigroup.
Constructing the theoretical image of open
systems requires a new understanding of how time parameters correspond
to the passage of time in the physical universe.
As a result, standard quantum mechanics
already predicts decoherence, without
invoking a reduced dynamics.
This suggests that the extrinsic arrow of time may
be the experimental signature of a more
fundamental, microphysical time asymmetry.

The practical result is a new framework for the treatment of decoherence, in which
states of open systems are represented by branching density operators rather
than by solutions of the decoherence master equation.
As an application, we have created a simple model matching
experimental results from Rabi oscillations experiments.
With the new formalism, we can conclude that a general yet puzzling
experimental result, known as Excitation Induced Dephasing,
is the measurable consequence of the indistinguishability
of separate, uncontrolled interactions between quantum systems and their
environment.

The new formalism is very promising for the study of quantum decoherence.
In forthcoming papers, we will show how the framework naturally extends to
statistical mechanics and the increase of quantum mechanical entropy.
And perhaps even more compelling, we will also demonstrate that this
same framework that works so well for decoherence is also very
effective when applied to scattering theory.

\end{document}